# Spin polarization separation of reflected light at Brewster angle

Yang Lv, Zefang Wang, Yu Jin, Mingtao Cao, Liang Han, Pei Zhang, Hongrong Li,
Hong Gao,* and Fuli Li

*Department of Applied Physics, Xi'an Jiaotong University, Xi'an 710049, China*
*\*Corresponding author: honggao@mail.xjtu.edu.cn*

A novel spin polarization separation of reflected light is observed, when a linearly polarized Gaussian beam impinges on an air-glass interface at Brewster angle. In the far-field zone, spins of photons are oppositely polarized in two regions along the direction perpendicular to incident plane. Spatial scale of this polarization is related to optical properties of dielectric and can be controlled by experimental configuration. We believe that this study benefits the manipulation of spins of photons and the development of methods for investigating optical properties of materials.

The spin Hall effect of light (SHEL) has been intensively investigated in both transmitted and reflected light [1-5]. The two spin components of reflected or transmitted beam receive opposite transverse displacements that are perpendicular to the incident plane. Recently, another type of spin dependent displacement called in-plane spin separation of light (IPSSL) also has been found [3]. The splitting of both SHEL and IPSSL is much smaller than the incident beam waist so that most areas of the two spin components overlap. But with the weak measurement technique, both SHEL and IPSSL are successfully observed in experiments [2-5]. In the particular case where a horizontally polarized Gaussian beam impinges on a dielectric interface, only SHEL occurs [3]. The transverse displacement of SHEL of reflected light deviates from zero dramatically when incident angle approaches Brewster angle [4,5]. And at exact Brewster angle, the displacement becomes dispersive. Actually, the reflected beam profile becomes distorted so that no valid displacement exists; if the incident beam is a Gaussian beam, it evolves to a superposition of two higher order modes [6,7]. As indicated in Ref. [6], the profile of the reflected light can be described by a superposition of a p-polarized $TEM_{10}$ and an s-polarized $TEM_{01}$ which share the same phase. The superposition of these two modes leads to a radial-like linear polarization with all polarization directions converging towards the beam center.

In this letter, we report a novel spin polarization separation of reflected light. We investigate the polarization distribution of the light reflected from an air-glass interface when a Gaussian beam incidents at the Brewster angle. We find that the reflected light is actually not purely linearly polarized but elliptically polarized. The reflected light close to beam center in two separate regions presents strong but opposite spin polarizations. The two regions are located along the direction perpendicular to incident plane. This new spin polarization separation is different from SHEL. First, the separation of the two spin polarization regions is much greater. Moreover, the spin polarization separation is not a spin dependent displacement but a spin dependent spatial accumulation.

Our experimental setup shown in Fig.1 is similar to those in Refs. [2-5] but has several modifications, a QWP is added so that polarization distribution can be measured. A Gaussian beam at 632.8nm is generated by a He-Ne laser and passes through a HWP, a Glan polarizer P1 and a lens L1. So, a horizontally linearly polarized focused Gaussian beam with a minimum beam waist of 14μm is generated and directed to glass prism at Brewster angle. The reflected beam is collimated by L2, whose focal plane would be placed at the waist of incident beam. The collimated reflected beam with an approximate diameter of 3mm passes through a QWP and P2 and is eventually captured by a charge-coupled device (CCD). The last three components construct a typical setup for measuring polarization distribution. By rotating the QWP to four specific positions, Stokes parameters can then be calculated from the four intensity distributions on CCD.

To accomplish the experiment, first we need to set the incident angle to Brewster angle and incident light polarization to perfect horizontal. This can be done by identifying the intensity profile after L2. The far-field intensity distribution is horizontally symmetric if the incident angle achieves exact Brewster angle. And it becomes vertically symmetric when the P1 is rotated to perfect horizontal. When all components are adjusted properly, we then retrieve the far-field polarization distribution of the reflected light. We choose 0, 30, 45 and 60 degree as the four positions QWP rotated to. And with the four pictures captured by CCD, we can calculate the distribution of Stokes parameters and then retrieve polarization distribution. We note that the polarization profile we measured is actually that of reflected beam transformed by L2. Further theoretical analysis shows that in the case where reflected light propagates freely, the far-field polarization distribution resembles that we measured after L2. Thus, the experimental results are indeed representative to the free propagation case.

We gather polarization information of reflected beam in terms of Stokes parameters. To better interpret these parameters, we introduce $\chi$ and $\theta$ defined as:

$$\chi = \frac{1}{2}\arcsin\left(S_3/\sqrt{S_1^2+S_2^2+S_3^2}\right), \quad (1)$$

$$\theta = \frac{1}{2}\arctan(S_2/S_1), \quad (2)$$

where $S_{1,2,3}$ are stokes parameters. $\chi$ is ellipticity angle, whose value of zero corresponds to linear polarization and value of ±45 degree corresponds to left-hand or right-hand circular polarization. $\theta$ is the angle between the major axis of the polarization ellipse and the horizontal ($x_O$)-axis.

Figure 2 (a) shows $\chi$ distribution on the observing plane. Photons with different spins are separated into two regions along vertical ($y_O$)-axis. And there are two peak points ($x_O = \pm x_O^{peak} \approx 0, y_O = \pm y_O^{peak}$) on $y_O$-axis where $\chi$ reaches almost 45 and -45 degree, respectively. This means that at these two points photons are detected at nearly pure $|+\rangle$ or $|-\rangle$ state. And these two points are as much as 1.1mm apart.

Figure 2 (b) shows $\theta$ distribution on the observing plane. The distribution shows a distinct difference with that one could expect from the conclusion of Ref.6. There are actually two rather than one odd points (or centers) where the contour lines of $\theta$ start from. These lines are curve near the odd points and become straight when far away from those odd points. These odd points are also located at $(0, \pm y_O^{peak})$.

The above new phenomenon of spin polarization separation could not be interpreted by current theoretical model of SHEL. Because at Brewster angle, some mathematical treatment become invalid; there is no valid displacement for p-polarized incident beam. Other studies suggest that the reflected light is a superposition of a horizontally polarized TEM$_{10}$ and a weaker vertically polarized TEM$_{01}$ [6]. However, this theoretical model still cannot explain what we observed. According to our analysis, this phenomenon is related closely to some subtle optical behaviors of glass. Since the reflection is very weak at Brewster angle, those behaviors become very crucial. Therefore, we conduct another experiment aiming at investigating the reflection behavior of glass at Brewster angle.

The experimental setup is similar to that of our first experiment. In addition, we remove L1 and L2 so that incident beam would be an approximately collimated light. Then the CCD is replaced with a photon detector (Thorlabs PDA36A-EC) which measures the reflective beam power after P2. First we remove the QWP and measure the power of reflected p-polarized component when a p-polarized beam incidents at Brewster angle. We find that about 4nW of p-polarized light leaks upon reflection at Brewster angle with about 2mW input power. The effective amplitude reflective coefficient we measured is $1.44(1.36,1.51)\times 10^{-3}$ at Brewster angle.

To determine the phase shift of reflected p-polarized light, we then add the QWP before P2 and rotate it to 45 degree. We slightly rotate P1 to let the reflected s-polarized light have the same amplitude with the p-polarized one. It is certain that the s-polarized light receives 180 degree phase shift upon reflection. If the p-polarized component receives a ±90 degree phase shift upon reflection, the reflected beam will be a perfect circularly polarized beam which would suffer a complete extinction when passing through QWP and then P2. Actually a nearly zero power output after P2 is detected in our experiment. Thus, we conclude that the reflected p-polarized component receives an approximate ±90 degree phase shift upon reflection. With the results of the second experiment, we find that we could describe the reflection leakage of p-polarized light and its phase shift by taking the imaginary part of refractive index of glass into the Fresnel equations. Therefore, we can calculate the effective imaginary part of refractive index of our prism and it is $(7.7\pm 0.4)\times 10^{-3}$.

In the following theoretical analysis of our first experiment, we would consider the refractive index of air as a real quantity $\tilde{n}_0 = n_1$ and that of glass as a complex quantity $\tilde{n}_2 = (n_2 + i\kappa)$.

Since incident and reflected beams contain distributions of wave-vectors, the Fresnel reflection coefficients and projections of different states of light vary with wave-vectors. This leads to a complicated profile of reflected light. After reflection, light evolves and is transformed by L2 and observed on the observing plane. The derivation is similar to that in Refs. [2,3]. Assuming the incident beam is a Gaussian beam, we obtain the state of reflected light observed on the observing plane:

$$|\varphi^{(O)}\rangle = h^{(O)}(x_O,y_O)|H\rangle + v^{(O)}(x_O,y_O)|V\rangle, \quad (3)$$

where

$$h^{(O)}(x_O,y_O) = A_0 \exp\left[-k^2\omega_o^2(x_O^2+y_O^2)/4f_2^2\right]$$
$$(if_2 h_0 + kh_1 x_O),$$

$$v^{(O)}(x_O,y_O) = A_0 \exp\left[-k^2\omega_o^2(x_O^2+y_O^2)/4f_2^2\right]$$
$$kv_1 y_O.$$

$k$ is length of wave-vector of incident light; $\omega_o$ is minimum beam waist of incident beam; $f_2$ is focal length of L2; $h_0$, $h_1$ and $v_1$ are real constants related to refractive indexes; $A_0$ is a constant related to $k$, $\omega_o$ and $f_2$. $|H\rangle$ and $|V\rangle$ are eigenstates in H-V basis.

In Eq. (3), we find an additional horizontally polarized fundamental mode component with a 90 degree phase shift, by taking the imaginary part of refractive index into account. The reflected light is actually a superposition of a horizontally polarized TEM$_{10}$, a vertically polarized TEM$_{01}$ and a horizontally polarized TEM$_{00}$ with a 90 degree phase shift. Using the result of the second experiment and other parameters in our first experiment, we extract $\chi$ and $\theta$ from Eq.(3) and plot insets of Fig.(2). The theoretical results show excellent agreement with experimental results.

To better understand the spin polarization separation, we express $|\varphi^{(O)}\rangle$ in spin basis as:

$$|\varphi^{(O)}\rangle = c^{(O)}(x_O,+y_O)|+\rangle + c^{(O)}(x_O,-y_O)|-\rangle, \quad (4)$$

where

$$c^{(O)}(x_O, y_O) = A_0 \exp\left[-k^2\omega_o^2\left(x_O^2 + y_O^2\right)/4f_2^2\right]$$
$$\left[h_1 x_O + iv_1\left(y_O + y_O^{peak}\right)\right],$$
$$y_O^{peak} = \left|\frac{f_2 h_0}{kv_1}\right| = \frac{n_2(n_1^2 + n_2^2)}{2n_1 n_2^3} f_2 \kappa.$$

Equation (4) shows that an effect of spin dependent flip along $y_O$-axis affects the two spin amplitudes of reflected light. For an ideal dielectric, $\kappa = 0$, amplitudes of spins are symmetric along $y_O$-axis, so the spin dependent flip alone does not induce spin polarization. However, for a realistic dielectric, when $\kappa > 0$, the symmetry of spin amplitudes along $y_O$-axis is broken. This asymmetry of spin amplitudes, together with the effect of spin dependent flip, finally induces the spin polarization of reflected light.

Moreover, we can see that the separation ($2y_O^{peak}$) of the two peak points, where $\chi$ reaches 45 and -45 degree, respectively, is proportional to $\kappa$ and is amplified by $f_2$. This result indicates that we can control the spin polarization separation by adjusting the focal length of L2. It may also provide a method for measuring the effective imaginary part of refractive index of dielectrics.

In conclusion, we have observed a novel spin polarization separation when a p-polarized Gaussian beam incidents air-glass interface at Brewster angle. Two regions with opposite spin polarization are separated along vertical direction. This separation can be controlled by experimental setup and is related to some optical properties of glass. This phenomenon is different from the well-know SHEL or IPSSL, which are induced by spin-orbit interaction. It can be well explained by taking account of imaginary part of refractive index. This study provides a new insight of profile of light reflected from a common and realistic glass surface at Brewster angle. We believe such phenomenon also provides a pathway for manipulating spins of photons and investigating optical characteristics of materials at surface.

We thank Dr. Dong Wei for very useful advice on this work. We acknowledge financial support from the National Natural Science Foundation of China (NSFC) under grants 11074198, 11004158, 11174233, and Special Prophase Project on the National Basic Research Program of China under grants 2011CB311807.

### References
1. M. Onoda, S. Murakami, and N. Nagaosa, Phys. Rev. Lett. **93**, 083901 (2004).
2. O. Hosten and P. Kwiat, Science **319**, 787 (2008).
3. Yi Qin, Yan Li, Xiaobo Feng, Yun-Feng Xiao, Hong Yang, and Qihuang Gong, Opt. Express **19**, 9636 (2011).
4. Y. Qin, Y. Li, H. Y. He, and Q. H. Gong, Opt. Lett. **34**, 2551 (2009).
5. Hailu Luo, Xinxing Zhou, Weixing Shu, Shuangchun Wen, and Dianyuan Fan, Phys. Rev. A **84**, 043806 (2011).
6. M. Merano, A. Aiello, M. P. van Exter, and J. P. Woerdman, Nat. Photonics **3**, 337 (2009).
7. O. V. Ivanov and D. I. Sementsov, Opt. Spectrosc. **92**, 419 (2002).

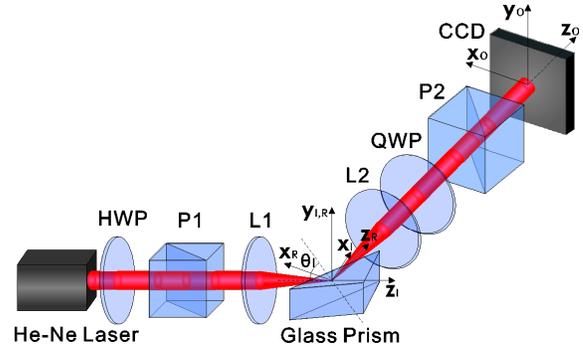

Fig. 1. Experimental setup for observing novel spin polarization separation of light reflected from air-glass interface. The He-Ne laser outputs a Gaussian beam at 632.8nm; HWP, half-wave plate for adjusting beam intensity; P1 and P2, Glan polarizers; L1 and L2, lenses with 44mm and 100mm effective focal lengths, respectively; Glass Prism, GCL-030105 manufactured by CDHC-OPTOELECTRONICS and made of BK7; QWP, quarter-wave plate, along with P2 and CCD (Lumenera Infinity 3-1), for gathering polarization distribution of reflected beam. $\mathbf{z}_I$ and $\mathbf{z}_{R,O}$ attaches to the central wave vector of incident and reflected beam respectively. $\mathbf{x}_{I,R,O}$ are parallel to the incident plane.

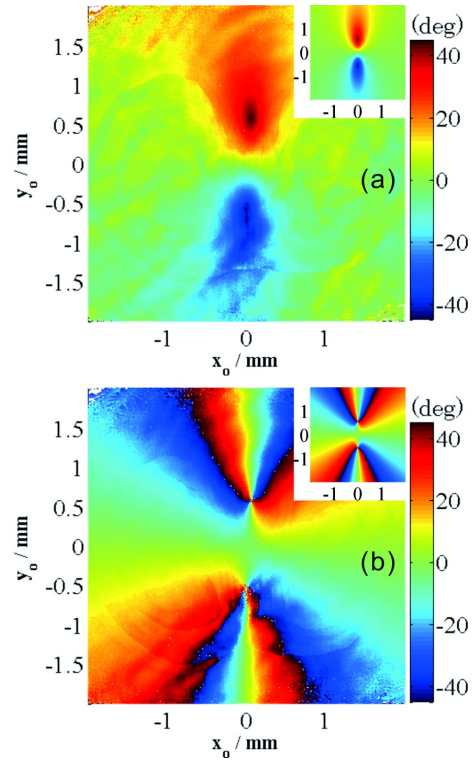

Fig. 2. $\chi$ (a) and $\theta$ (b) distributions of reflected light observed on the observing plane. To eliminate spatial noise of CCD chip, a mean filtering with 5×5 kernel is performed for every image before the images are used to retrieve polarization distribution. The original sample size is 800×800 pixel, with one pixel size of 6.45×6.45 μm². Insets show theoretical prediction of $\chi$ and $\theta$ distributions when using data acquired through the second experiment and other parameters in our first experiment.

Information page

1. M. Onoda, S. Murakami, and N. Nagaosa, "Hall effect of light," Phys. Rev. Lett. 93(8), 083901 (2004).
2. Hosten and P. Kwiat, "Observation of the spin hall effect of light via weak measurements," Science 319(5864), 787–790 (2008).
3. Yi Qin, Yan Li, Xiaobo Feng, Yun-Feng Xiao, Hong Yang, and Qihuang Gong, "Observation of the in-plane spin separation of light," Opt. Express 19(10), 9636–9645 (2011).
4. Y. Qin, Y. Li, H. Y. He, and Q. H. Gong, "Measurement of spin Hall effect of reflected light," Opt. Lett. 34(17), 2551–2553 (2009).
5. Hailu Luo, Xinxing Zhou, Weixing Shu, Shuangchun Wen, and Dianyuan Fan, "Enhanced and switchable spin Hall effect of light near the Brewster angle on reflection," Phys. Rev. A 84(4), 043806 (2011).
6. M. Merano, A. Aiello, M. P. van Exter, and J. P. Woerdman, "Observing angular deviations in the specular reflection of a light beam," Nat. Photonics 3(6), 337–340 (2009).
7. V. Ivanov and D. I. Sementsov, "Transformation of a Gaussian Light Beam Reflected in the Vicinity of the Brewster Angle," Opt. Spectrosc. 92(3), 419–424 (2002).